\documentclass[12pt]{article}
\usepackage{cprotect}
\usepackage{enumitem}
\usepackage{layout}
\usepackage{caption}
\usepackage{subcaption}
\usepackage{amssymb}
\usepackage{microtype}
\usepackage{diagbox}
\usepackage{graphicx}
\usepackage{booktabs}
\usepackage[table,xcdraw]{xcolor}

\usepackage{stylesheet} 
\usepackage{authblk}
\usepackage{hyperref}

\title{\vspace{-1.5cm}Conformal Anomaly Detection on Spatio-Temporal Observations with Missing Data}
\author{\vspace{-0.3cm}Chen Xu\footnote{cxu310@gatech.edu}
\qquad Yao Xie\footnote{yao.xie@isye.gatech.edu}}
\affil{School of Industrial
and Systems Engineering \\ Georgia Institute of Technology
}
\date{\vspace{-6ex}}
       
\begin{document}
\maketitle
\vspace{-0.35cm}
\begin{abstract}
    We develop a distribution-free, unsupervised anomaly detection method called \ECAD, which wraps around any regression algorithm and sequentially detects anomalies. Rooted in conformal prediction, \ECAD \ does not require data exchangeability but approximately controls the Type-I error when data are normal. Computationally, it involves no data-splitting and efficiently trains ensemble predictors to increase statistical power. We demonstrate the superior performance of \ECAD \ on detecting anomalous spatio-temporal traffic flow.
\end{abstract}

\section{Introduction}

Detecting whether new observations are anomalies or outliers is critical in many applications, such as network cyber-intrusions \citep{AD_app_cyber}, healthcare \citep{AD_app_health}, video surveillance \citep{AD_app_video}, etc. There have been many parametric detection methods \citep{AD_parametric2,AD_parametric1,AD_parametric3}, but parameter estimation and model selection can limit their applicability. Meanwhile, nonparametric methods \citep{AD_nonpara1,AD_nonpara2} avoid distributional assumptions but may not provide theoretical guarantees, such as 
Type-I error control and power analyses. It is also unclear how well they tackle data with dependency and missing observations. In general, theoretically sound distribution-free anomaly detection for such complex data needs to studied better.

Conformal prediction (CP) is a natural and general scheme that uses past observations to quantify where normal observations lie \citep{conformaltutorial}. The essential idea assigns ``conformity scores'' to training and test observations. Data with small scores are likely anomalies. In particular, methods following this scheme are guaranteed to control the Type-I error, as long as observations are exchangeable (e.g., iid). Power analyses are possible, depending on how scores are assigned and compared. Many works under this logic have been proven successful in uncertainty quantification for regression \citep{CPquantile,j+ab} and classification \citep{Candes_classification,MJ_classification}. 

CP methods have also been adopted for anomaly detection. \cite{conformal_knn,conformal_kernel} combine KNN and Mahalanobis distance or uses kernel functions to assign anomaly scores locally for univariate time-series data. However, these methods only use a subset of training data to train the detector, with the rest being calibration data, likely reducing power. Moreover, although the theoretical guarantee provided in that work is a corollary from \citep{vovk_cond_val}, how time-series therein are exchangeable data is left unanswered. \cite{conformal_supper} proposes a frequency-based conformity score for anomaly detection, but the method is limited to iid and exchangeable data. To the best of our knowledge, our method is the first conformal anomaly detector that applies to spatial-temporal observations and provides approximately valid theoretical guarantees.

In this work, we are primarily interested in detecting real-valued anomalous spatial-temporal observations that arrive in sequence, assuming missing values are present in training data. To this end, we build an Ensemble Conformal Anomaly Detection (\ECAD), based on the method introduced in \citep{EnbPI}. It uses residuals from ensemble regression models as anomaly scores and makes detection through local comparisons, with approximate marginal coverage guarantee for time-series observations under mild assumptions. 
The rest work is organized as follows: Section \ref{setup} sets up the detection problem, introduces \ECAD \ and its novelties. Section \ref{ECAD} explains \ECAD \ procedures in detail. Section \ref{expr} applies \ECAD \ on anomalous traffic flow detection. Section \ref{conclu} concludes the work and discusses extension ideas.

\section{Problem Setup and Background}\label{setup}

Given a network graph $G=(N,E)$, where $|N|=K$, each node/sensor $k\in \{1,\ldots,K\}$ records a univariate time-series $Y_{tk}\in \mathbb R, t=1,2\ldots$ In particular, we assume no knowledge of $E$. Instead, assume there is a common data-generating model $f:\mathbb R^d\rightarrow \mathbb R$ for all sensors: 
\begin{equation}\label{model}
    Y_{tk}=f(X_{tk})+\epsilon_{tk},
\end{equation}
where $X_{tk}\in \mathbb R^d$ is an \textit{unknown} feature vector (e.g., may be comprised of past and neighboring observations), and the errors $\epsilon_{tk}\sim \mathcal{D}$ are independent over sensors $k$ but possibly dependent over time $t$ at each $k$. We consider anomalies as observations $Y_{tk}$ that violate (\ref{model}), such as having extremely large/small values. In reality, defining features $X_{tk}$ often requires domain expertise and knowledge of the network topology encoded in edges $E$. 

We remark that assumptions such as sensor-invariant $f$ and identically distributed errors $\epsilon_{tk}$ at each sensor are purely computational, as we thus only need to train \ECAD \ once on data from all sensors to approximate $f$. In general, (\ref{model}) can be relaxed so that at each sensor, $f$ becomes $f_k:\mathbb R^{d_k}\rightarrow \mathbb R$ and errors $\boldsymbol \epsilon_k:=\{\epsilon_{tk}\}_{t\geq 1}$ follow the distribution $\mathcal{D}_k$. However, we will see that such flexibility requires training \ECAD \ $K$ times, one for each sensor; doing so is not scalable to problems with large $K$ (e.g., social networks) and computationally expensive regression algorithms (e.g., deep neural networks).

Prior to the anomaly detection, we assume access to an \textit{incomplete} training data matrix $\boldsymbol Y=\{Y_{tk}\}\in \mathbb R^{TK}$. The data are missing in the sense that for each $k=1,\ldots,K$, only a subset $T_k \subset T$ of entries in $\boldsymbol Y_t=[Y_{1k},\ldots,Y_{Tk}]$ are recorded in $\boldsymbol Y$. The rest are marked unobserved. Moreover, $T_{k_i} \neq T_{k_j}$ as long as $k_i \neq k_j$, as data can be missing at different times for different sensors. For traffic flow detection in Section \ref{expr} (Experiments), we thus face a situation in which sensor fail to record traffic flows asynchronously.

Our proposed \ECAD \ is based on the recent \EnbPI \ algorithm in \citep{EnbPI}. In short, it first fits ensemble predictors using a regression model $\mathcal{A}$ and then outputs $p$-values via locally ranking the test residuals against ``leave-one-out'' (LOO) training residuals. Observations with small $p$-values are identified as anomalies. Detailed descriptions are in Section \ref{ECAD}. 

\section{Method: Ensemble Conformal Anomaly Detection}\label{ECAD}

The \ECAD \ has two primary phases, the training phase that trains the ensemble anomaly detector and the detection phase that detects anomalies sequentially. We only consider data with single subscripts below for notation simplicity but will clarify how to apply \ECAD \ to spatio-temporal data.

\vspace{.05in}
\noindent\textbf{Training Phase in \ECAD.}

\noindent   (1) Fix a regression algorithm $\mathcal{A}$, which trains on $N$ data points $\{(X_i,Y_i)\}_{i=1}^N$ and outputs a non-conformity mapping (NCM) $\mathcal{N}:\mathbb R^d \times \R \rightarrow \mathbb R$. The input to $\mathcal{N}$ is any new data point $(X_j,Y_j)$ and the output is the anomaly score $\hat s_j:=|Y_j-\hat{f}(X_j)|$, which is the residual from the fitted model. 
    
    
    \noindent  (2) Assuming $T'$ out of $T$ training data are non-missing, we then train $B$ bootstrap NCM. Formally, for each $b=\{1,\ldots,B\}$, define $\mathcal{N}_b:=\mathcal{A}(\{(X_j,Y_j)\}_{j=b_1}^{b_{T'}})$ as the $b$-th NCM, where $\mathcal{A}$ trains on $T'$ data points sampled with replacement from $T'$ non-missing data. One may also impute the training data and then sample from the full data with imputed values, under which $T'$ becomes $T$.
    
    
    \noindent (3) For the $i$-th non-missing training observation, we aggregate NCM from (2) in a LOO fashion via a function $\phi$ to compute the aggregated anomaly score $\hat s_i^{\phi}$ for $(X_i,Y_i)$. Examples of $\phi$ include mean, median, or trimmed mean. Formally, for each $i\in \{1,\ldots, T'\}$, calculate $\hat s_i^{\phi}:=|Y_i-\phi(\{\hat{f}_b(X_i): i\notin \{b_1,\ldots, b_{T'}\}\}_{b=1}^B)|$. Note that $\hat{f}_b$ is the bootstrap predictor trained for $\mathcal{N}_b$. We will then use $\{\hat s_i^{\phi}\}_{i=1}^{T'}$ as the initial set of scores to compute detection thresholds. Furthermore, let $\hat f^{\phi}_{-i}:=\phi(\{\hat{f}_b: i\notin \{b_1,\ldots, b_{T'}\}\}_{b=1}^B)$ be the $i$-th LOO ensemble predictor, where aggregation is applied pointwise.

\vspace{.05in}
\noindent \textbf{Detection Phase in \ECAD.}

 \noindent (1) To detect whether the test datum $(X_t,Y_t), t>T$ is an anomaly, calculate \[\hat s^{\phi}_t:=|Y_t-(1-\alpha)\text{ quantile of }\{\hat{f}^{\phi}_{-i}(X_t)\}_{i=1}^{T'}|,\] where $\alpha$ is the user-defined significance level. Then, compute the $p$-value $p_t:={T'}^{-1}\sum_{i=1}^{T'} \textbf{1}(\hat s^{\phi}_i \geq \hat s^{\phi}_t)$, where $\textbf{1}(\cdot)$ is the indicator function. Lastly, call $x_t$ an anomaly if $p_t\leq \alpha$. 
    
    \noindent (2) Slide the past sequence of $T'$ anomaly scores ahead, so $s_1$ is dropped and $s_t$ appended. Reset indices afterwards. Exist this phase if no more detection needs to be made; otherwise, return to (1).

We briefly elaborate on several theoretical and computational benefits of \ECAD \ for anomaly detection, which are inherited from \EnbPI \ \citep{EnbPI}. Theoretically, it approximately controls the Type-I error but does not assume the exchangeability of observations.\footnote{See \citep[Theorem 1]{EnbPI}.} The only data assumption is on the dependency of \textit{errors} $\boldsymbol \epsilon_k$ at each sensor.\footnote{See \citep[Assumption 1 and Corollary 1--3]{EnbPI}.} Meanwhile, it requires consistency of predictors $\hat{f}$ to the data model $f$, which holds for many regression models, including neural networks. Computationally, it does not involve data-splitting as in the split/inductive conformal \citep{inductCP}. Lastly, it merely requires the computational cost of fitting one ensemble predictor but outputs $T$ LOO ensemble predictors for each sensor, ensuring computational efficiency and increasing power.

We additionally comment on the computation and practical implementation of \ECAD. 1). $B$ between 20 to 30 is sufficient for good performance; due to concentration inequality, each LOO ensemble predictor likely aggregates over a balanced number of bootstrap models in (3) of the training phase. 2). The most computationally expensive part happens in (2) of the training phase, especially when $\mathcal{A}$ is a deep neural network. However, because the neural network trains on bootstrap samples of the original data and \cite{EnbPI} demonstrates good performance even for problems with few training data, (2) is still computationally affordable in practice. 3). If there are change points during test time that alter $f$ in (\ref{model}), one may want to repeat the training phase for better performance so that LOO ensemble predictors approximate post-change $f$ better.

\section{Experiment: Detecting Anomalous Traffic Flows}\label{expr}

\noindent \textbf{Dataset. }We first describe the complete data and then outline how anomalies are defined for each sensor, as well as how missing entries are selected. The dataset contains observations from 20 sensors, which record hourly traffic flow every day in 2020. Each sensor is provided a pair of scaled latitude and longitude coordinates so that the $l_2$-norm distance between any two sensors is within $[0,1]$. The average pairwise distance between all sensors is 0.607. Data in the first six months are reserved for training, and the rest are for testing the performance of detectors. Overall, the training data contain $K=20$ sensors and $T=8783$ univariate flow observations at each sensor, with unknown interactions among sensors. Then for each $k \in \{1,\ldots,20\}$ and $t\in \{1,\ldots,17520\}$ (full hourly data in 2020), define $Y_{tk}$ as an anomaly if 
\begin{equation}\label{anomaly_def}
    Y_{tk} \geq q_{1-\alpha}(Y^{-d}_{t,N_k}) \text{ or } Y_{tk} \leq q_{\alpha}(Y^{-d}_{t,N_k}),
\end{equation}
where $N_k$ contains sensors closest to $k$ (including itself) and $Y^{-d}_{t,N_k}$ contains past $d$ hourly flows from sensors in $N_k$. In words, $Y_{tk}$ is defined as an anomaly if it is too large/small comparing to the past and nearby flows. We let $q_{\alpha}(\cdot)$ be the $\alpha$ percentile of its input vector and choose $\alpha=0.01$, $d=3$, and $|N_k|=4$, all three of which are unknown to \ECAD.

Lastly, we randomly select 40$\%$ training data to be missing for each sensor $k$, so that $|T_k|=0.6T$ and with high probability, $T_k \neq T_{k'}$ if $k \neq k'$.

\vspace{.05in}
\noindent \textbf{\ECAD \ and Competing Methods.} We first use the \Verb|IterativeImputer| from the Python \Verb|sklearn| package to impute the missing entries in $\boldsymbol Y$ column-wise. Specifically, we iteratively impute columns of the data matrix by regressing the $k$-th column $\boldsymbol Y_k$ on the rest columns $\boldsymbol Y_{-k}$. The missing values are imputed as predictions from the fitted regression model. Note that we do not remove observations labeled as anomalous from the incomplete training data matrix because features $X_{tk}$ likely depends on both normal and anomalous data, and definition (\ref{anomaly_def}) does not assume only normal observations are present in $Y^{-d}_{t,N_k}$ (although (\ref{anomaly_def}) is unknown to us). Notation-wise, we still use the symbol $\boldsymbol Y$ to denote the imputed data matrix.

We then apply \ECAD \ to the imputed $\boldsymbol Y$ as follows. Firstly, define $X_{tk}:=Y^{-m}_{t,\hat{N}_k} \in \mathbb R^{m|\hat{N}_k|}$ as the past $m$ hourly flows from $\hat{N}_k$ closest sensors to $k$. We let $m=|\hat{N}_k|=5$ for all $k$. Secondly, we choose $\mathcal{A}$ as the RNN with 2 LSTM layers, each with 100 hidden sensors and the Tanh activation function. The dense layer uses the ReLU activation function. The full network is optimized using Adam. We believe RNN is useful because the current flow depends strongly on past and neighboring flow, so that $f$ in (\ref{model}) can be better approximated using RNN. Thirdly, given the imputed data matrix $\boldsymbol Y \in \mathbb R^{TK}$, we subsample over $T$ for each bootstrap RNN model, so that the $b$-th bootstrap model has access to all rows $\boldsymbol Y_t \in \mathbb R^{K}, t \in \{b_1,\ldots, b_T\}$. Lastly, during detection, we only compare $\hat s^{\phi}_{tk}$ against anomaly scores $\hat s^{\phi}_{t'k'}$ if $t-m\leq t'\leq t-1$ or $k\in \hat{N}_k$, because residuals too far away from time $t$ or sensor $k$ may mislead detection. The significance level $\alpha$ is fixed at 0.05.

We consider 10 competing anomaly detectors, 4 of which are unsupervised methods and 4 of which are supervised methods. All competitors have access to the same training data as \ECAD \ does, where we additionally provide training labels for supervised methods. The first two are \ECAD \ under ridge regression and the random guess classifier (RGuess). RGuess is a fair coin that predicts flows as normal/anomalous with equal probability. Then, the four unsupervised detectors are: The Histogram-based Outlier Detection (HBOS) \citep{HBOS}, Isolation Forest (IForest) \citep{IForest}, One-class Support Vector Machine (OCSVM) \citep{OCSVM}, and the Principal Component Analysis (PCA) \citep{PCA}. They are all implemented in the \Verb|PyOD| package for python \citep{PYOD}. Lastly, the four supervised detectors are: Gradient Boosting Classifier (GBoosting) \citep{GBoosting}, Multi-layer Perceptron Classifier (MLP) \citep{MLP}, Logistic Regression (Logistic) \citep{Logistic}, and Random Forest Classifier (RF) \citep{RF}. These four all come from the \Verb|sklearn| package. All ten competitors except \ECAD \ under ridge and RGuess operate under their default parameter setting. 

\vspace{.05in}
\noindent \textbf{Results.} Our primary aim is to demonstrate that \ECAD \ with ensemble RNN achieves the highest $F_1$ score in many cases. A high $F_1$ score indicates that the detector identifies many true positives and keeps both false negatives and false positives low. Note that the maximum expected $F_1$ score for RGuess is $2q/(q+1)$ upon predicting all observations as anomalous (recall is 1 and precision is $q$), so we operate it in this way onward; keep in mind that RGuess is thus a biased detector.

Table \ref{tab:f1_tab} shows the performance of all 11 anomaly detectors. Under access to the same set of data, \ECAD \ RNN outperforms the other methods by a large margin (e.g., 0.2 $F_1$ score or more). Although it seems that $F_1$ scores by \ECAD \ RNN positively and strongly correlate with anomaly fractions, such phenomena do not occur on all 20 sensors. At some sensors where \ECAD \ RNN performs slightly worse than competitors, anomaly fractions can be higher than 50\%. However, we still think anomaly fraction crucially affects the performance of \ECAD \ RNN for the following reason. Sensors with high and low anomaly fractions likely have different data-generating models. By fitting \ECAD \ RNN only once on data from all sensors, we then likely capture less true positives at sensors with low anomaly fractions, simply because fewer anomalies are present. To remedy this, one may fit \ECAD \ RNN separately on each sensor or a group of sensors with similar anomaly fractions and/or geographic features to better approximate the unknown data-generating models. On the other hand, how users define $X_{tk}$ and imputation procedures also affect the performances, and we will explore these questions in the future.

\begin{figure}[htbp]
    \centering
    \includegraphics[width=\linewidth]{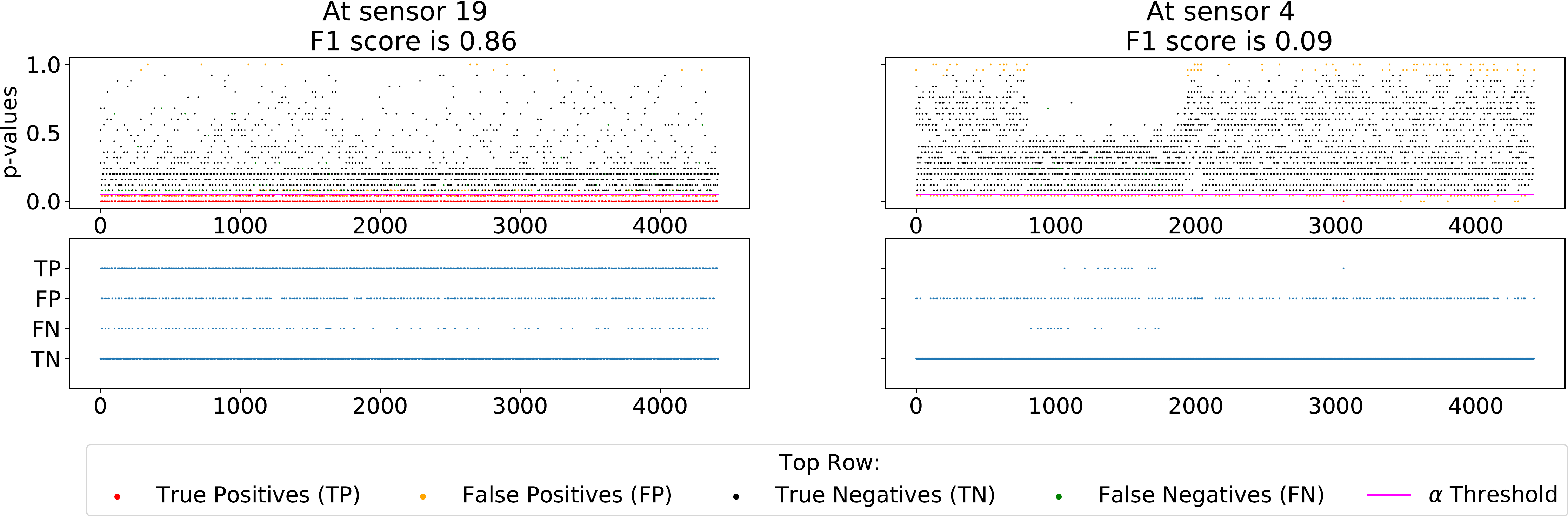}
    \vspace{-0.3cm}
    \cprotect\caption{Performance of \ECAD \ RNN on sensors at two extremes: sensor 19 on the left has roughly 32\% anomalies, so that most predictions are TP and TN. In contrast, sensor 4 on the right has only 0.66\% anomalies, so that FP and TN dominate predictions.}
    \label{fig:pval}
\end{figure}

In addition, Figure \ref{fig:pval} shows the trajectory of \ECAD \ detections at sensors 19 and 4, which helps one better understand the behavior of \ECAD \ RNN at different sensors with dramatically different anomaly fractions. Recall from Table \ref{tab:f1_tab} that \ECAD \ RNN performs the best among all methods at sensor 19. Although not shown earlier, \ECAD \ RNN barely identifies any anomalies at sensor 4. Based on definition (\ref{anomaly_def}), only 0.66\% data at sensor 4 are anomalous when sensor 19 has 32\% anomalous hourly flows. Hence, data at sensor 4 are much more imbalanced. As a result, we observe in Figure \ref{fig:pval} that \ECAD \ RNN can accurately capture most of the anomalies (i.e., positive events) at sensor 19, resulting in a balanced number of false and true positives, with very few false negatives. On the other hand, it performs poorly at the other sensor since there are nearly no anomalies and the data-generating model is likely very different from the rest. To improve the performance, one should fit \ECAD \ RNN on data from sensor 4 separately.

\section{Conclusion}\label{conclu}
Our work proposes an unsupervised Ensemble Conformal Anomaly Detector for spatio-temporal continuous data with missing entries. Besides being computationally efficient and easy to implement, it can achieve approximately valid marginal coverage without requiring exchangeability in data. In particular, it is scalable to large-scale problems with missing observations. Moreover, it shows superior empirical performance on anomalous traffic flow detection than competing methods. 

There multiple empirical and theoretical extensions. Empirically, 1) Adapt \ECAD \ for classification \citep{Candes_classification,MJ_classification}, such as detecting images under adversarial attacks \citep{Goodfellow_adversarial,Samangouei_adversarial,Pang2019_adversarial}. 2) Adopt other deep learning models as baselines $\mathcal{A}$ in \ECAD, which may be more suitable for anomaly detection \citep{review_DL_AD}. 3) Address poor performances when data have few anomalies. 4) Better define features $X_{tk}$ and impute missing entries, by taking into account the graph topology (e.g., interactions among sensors). Theoretically, 1) Explore how alternative assumptions on the data-generating model and errors affect marginal Type-I error control. 2) Use multiple testing corrections to lower bound precision \citep{contextual-fdr,LORD++,candes_anomaly}. The crucial step is examining $p$-value dependency in \ECAD. 3) Move towards conditional coverage guarantee by imposing additional assumptions, as distribution-free conditional validity is impossible  \citep{Lei_condcoverage}. 
\begin{table}[htbp]
\begin{center}
    \begin{tabular}{lllll}
\toprule
Sensor (Anomaly Fraction)             & 19 (32\%)   & 14 (38\%)   & 5 (31\%)    & 3 (24\%) \\
\bottomrule
\end{tabular}
\vspace{-0.4cm}
\end{center}
\begin{subtable}{\textwidth}
\resizebox{\textwidth}{!}{%
\begin{tabular}{c|cllllllllll}
\toprule
& \multicolumn{10}{c}{$F_1$ Score}\\
\backslashbox{Sensor}{Detector} &
RGuess & 
{\color[HTML]{000000} \ECAD \ RNN} &
{\color[HTML]{000000} \ECAD \ Ridge} &  {\color[HTML]{000000} HBOS} & {\color[HTML]{000000} IForest} & {\color[HTML]{000000} OCSVM} & {\color[HTML]{000000} PCA} & {\color[HTML]{000000} GBoosting} & {\color[HTML]{000000} MLP} & {\color[HTML]{000000} Logistic} & {\color[HTML]{000000} RF} \\
\midrule
19                                              & 0.48              & \textbf{0.86}                   & \textit{0.51} & 0.03                        & 0.11                           & 0.48                         & 0.31                       & 0.26                             & 0.20                       & 0.21                            & 0.24                      \\
14                                              & 0.55              & \textbf{0.84}                   & \textit{0.61} & 0.02                        & 0.00                           & 0.57                         & 0.00                       & 0.32                             & 0.30                       & 0.43                            & 0.28                      \\
5                                               & 0.48              & \textbf{0.84}                   & \textit{0.49} & 0.05                        & 0.04                           & 0.47                         & 0.02                       & 0.29                             & 0.33                       & 0.29                            & 0.33                      \\
3                                               & 0.39              & \textbf{0.81}                   & \textit{0.61} & 0.07                        & 0.00                           & 0.40                         & 0.02                       & 0.04                             & 0.05                       & 0.02                            & 0.04                      \\
\bottomrule
\end{tabular}%
}
\end{subtable}
\begin{subtable}{\textwidth}
\resizebox{\textwidth}{!}{%
\begin{tabular}{c|cllllllllll}
\toprule
& \multicolumn{10}{c}{Precision}\\
\backslashbox{Sensor}{Detector} &
RGuess & 
{\color[HTML]{000000} \ECAD \ RNN} &
{\color[HTML]{000000} \ECAD \ Ridge} &  {\color[HTML]{000000} HBOS} & {\color[HTML]{000000} IForest} & {\color[HTML]{000000} OCSVM} & {\color[HTML]{000000} PCA} & {\color[HTML]{000000} GBoosting} & {\color[HTML]{000000} MLP} & {\color[HTML]{000000} Logistic} & {\color[HTML]{000000} RF} \\
\midrule
 19                                              & 0.32              & \textbf{0.79}                   & \textit{0.52} & 0.10                        & 0.61                           & 0.31                         & 0.87                       & 0.24                             & 0.28                       & 0.23                            & 0.23                      \\
14                                              & 0.38              & \textbf{0.77}                   & \textit{0.67} & 0.10                        & 0.05                           & 0.40                         & 0.06                       & 0.37                             & 0.45                       & 0.40                            & 0.33                      \\
5                                               & 0.31              & \textbf{0.80}                   & \textit{0.58} & 0.06                        & 0.06                           & 0.31                         & 0.03                       & 0.26                             & 0.43                       & 0.27                            & 0.33                      \\
3                                               & 0.24              & \textbf{0.72}                   & \textit{0.59} & 0.17                        & 0.03                           & 0.25                         & 0.16                       & 0.04                             & 0.06                       & 0.03                            & 0.05                      \\
\bottomrule
\end{tabular}%
}
\end{subtable}
\begin{subtable}{\textwidth}
\resizebox{\textwidth}{!}{%
\begin{tabular}{c|cllllllllll}
\toprule
& \multicolumn{10}{c}{Recall}\\
\backslashbox{Sensor}{Detector} &
RGuess & 
{\color[HTML]{000000} \ECAD \ RNN} &
{\color[HTML]{000000} \ECAD \ Ridge} &  {\color[HTML]{000000} HBOS} & {\color[HTML]{000000} IForest} & {\color[HTML]{000000} OCSVM} & {\color[HTML]{000000} PCA} & {\color[HTML]{000000} GBoosting} & {\color[HTML]{000000} MLP} & {\color[HTML]{000000} Logistic} & {\color[HTML]{000000} RF} \\
\midrule
 19                                              & \textbf{1.0}                          & \textit{0.93}                   & 0.50       & 0.01                        & 0.06                           & 1.0                          & 0.19                       & 0.29                             & 0.15                       & 0.20                            & 0.25                      \\
14                                              & \textbf{1.0}                          & \textit{0.91}                   & 0.56       & 0.01                        & 0.00                           & 1.0                          & 0.00                       & 0.28                             & 0.22                       & 0.46                            & 0.24                      \\
5                                               & \textbf{1.0}                          & \textit{0.88}                   & 0.43       & 0.04                        & 0.03                           & 1.0                          & 0.01                       & 0.32                             & 0.27                       & 0.30                            & 0.33                      \\
3                                               & \textbf{1.0}                          & \textit{0.92}                   & 0.62       & 0.04                        & 0.00                           & 1.0                          & 0.01                       & 0.03                             & 0.04                       & 0.02                            & 0.04                      \\
\bottomrule
\end{tabular}%
}
\end{subtable}
\vspace{-0.3cm}
\cprotect\caption{$F_1$ scores, Precision, and Recall by 11 anomaly detectors on selected sensors. The scores are sorted in descending order by the $F_1$ score of \ECAD \ RNN, our desired method. Bold and italicized cells indicate the highest and second highest scores. In terms of $F_1$ scores, \ECAD \ RNN yields superior performance on this task.}
\label{tab:f1_tab}
\end{table}

\clearpage
\printbibliography

\end{document}